\documentclass[10pt, conference, compsocconf]{IEEEtran}
\usepackage{graphicx}
\usepackage{acronym}
\usepackage{subfigure}
\usepackage{booktabs}
\usepackage{algorithm}
\usepackage{algcompatible}
\usepackage{url}
\usepackage{amstext}
\usepackage[utf8]{inputenc}
\usepackage{amsmath,amssymb}

\acrodef{P2P}{Peer-to-Peer}
\acrodef{TTL}{Time-To-Live}

\begin{document}
\title{Publish-Subscribe Systems via Gossip:\\ a Study based on Complex Networks}

\author{
Stefano Ferretti\\
Department of Computer Science, University of Bologna\\
Bologna, Italy\\
sferrett@cs.unibo.it
}


\maketitle              

\begin{abstract}
This paper analyzes the adoption of unstructured P2P overlay networks to build publish-subscribe systems. We consider a very simple distributed communication protocol, ba\-sed on gossip and on the local knowledge each node has about subscriptions made by its neighbours. In particular, upon reception (or generation) of a novel event, a node sends it to those neighbours whose subscriptions match that event. Moreover, the node gossips the event to its ``non-interested'' neighbours, so that the event can be spread through the overlay. A mathematical analysis is provided to estimate the number of nodes receiving the event, based on the network topology, the amount of subscribers and the gossip probability. These outcomes are compared to those obtained via simulation. Results show even when the amount of subscribers represents a very small (yet non-negligible) portion of network nodes, by tuning the gossip probability the event can percolate through the overlay. Hence, the use of unstructured networks. coupled with simple dissemination protocols, represents a viable approach to build peer-to-peer publish-subscribe applications.
\end{abstract}
%


\section{Introduction}

Publish-subscribe is a distributed paradigm that gained a lot of attention in the last years. Today, it is widely used in several large-scale distributed applications, such as checking stock exchange quotations, information dissemination in social networks, order-processing systems, targeted advertising, multiplayer online games, decentralized business process execution, workflow management, business and system monitoring, discovery and general news dissemination \cite{Cheung:2010}.
The interesting feature of a publish-subscribe system is that it allows nodes to communicate asynchronously in a loosely and decoupled manner. This property gives systems higher modularity as well as easier maintainability. 
In a publish-subscribe system, there are nodes which are interested in receiving some type of contents. They are referred as \textit{subscribers}; in fact, to declare their interests they subscribe to these contents. \textit{Publishers} are those actors who produce information. 
Loose-coupling is achieved since producers do not have information on the identity and number of subscribers, as well as consumers subscribe to specific information without knowing the identity and number of possible publishers. 
Usually, novel contents published and sent to subscribers are referred as ``events'' \cite{Ahullo:2008}. 

Publish-subscribe systems can be implemented by resorting either to centralized or distributed solutions \cite{Carzaniga:2000}. Centralized solutions were the first to be implemented and, as for all centralized approaches, they have the advantage that the central server retains a global and up-to-date image of the system \cite{Eugster:2003,Fabret:2001,TamAJ03}. 
As usual, major disadvantages are the lack of scalability and fault-tolerance. 

On the other hand, several distributed publish-subscribe systems exist. 
The more interesting approaches are those based on Distributed Hash Tables (DHTs) \cite{Baldoni:2005,Shvartzshnaider:2010,Stoica:2003}. 
In few words, each node in the DHT is responsible for managing subscriptions/publications related to a given topic. Hence, each novel publication passes through the corresponding node in the DHT, which in turn triggers the dissemination of the published event to the appropriate subscribers.
These approaches result quite scalable and provide mechanisms to cope (up to some extent) with the arrival, departure and failure of nodes. However, these solutions impose constraints both on the overlay topology and on content placement in the overlay, so as to enable efficient discovery of data.
Contents must be usually classified into a fixed number of topics, so that they can be mapped 
into nodes in the DHT.
Moreover, this solution introduces an additional overhead to construct and maintain the overlay.
It has been also observed that these distributed solutions may lead to uneven load distribution, due to different
densities of contents and interests by end-users, which may imply that certain nodes are subject to more subscriptions/publications to handle \cite{Cheung:2010}.

A very different type of solutions relates to the use of unstructured \ac{P2P}~overlay networks \cite{Terpstra:2007,sub2sub,Wong:2008}. In an unstructured \ac{P2P}~overlay, links among nodes are established arbitrarily. 
They are particularly simple to build and manage, with little maintenance costs, yet at the price of a non-optimal organization of the overlay \cite{Lin:2009}. 
Peers locally manage their connections to build some general desired topology. Such a selected topology may vary depending on the characteristics the system should have. For instance, choosing a uniform graph where nodes have all the same degree (i.e.~number of connected nodes) might be useful to balance the load at peers for the distributed communication. Conversely, scale-free networks might be selected when the overlay needs to be robust and with a reduced network diameter \cite{newmanHandbook}. 
Whatever its form, in an unstructured overlay the links among peers do not depend on the contents being disseminated through it \cite{EberspacherS05a}.
Unstructured overlays are quite useful when the number of nodes is very high, with very frequent topology changes and churns, i.e.~high number of nodes joining and leaving the system.

Publish-subscribe systems can be built on top of unstructured networks by adopting either a gossip-based communication protocol, or some more sophisticated algorithm to route messages in the overlay. Contents can be replicated or not, as well as queries. In any case, we might sum up that these systems can be effectively employed when: 
i) the number of nodes is very high and dynamic, with high churn rates; 
ii) there is a high number of publications to handle; 
iii) there is a high number of subscribers to a given type of contents and hence usually an event must be propagated to reach a non-negligible portion of nodes in the overlay (the event must percolate through the network).

In this work, we study if a general \ac{P2P}~publish-subscribe system can be implemented on top of unstructured overlay networks. In particular, to distribute events through the unstructured overlay, we consider a simple dissemination protocol which is based only on local knowledge among peers in the network. 
Each node knows its own subscriptions and those of its neighbours only. Hence, each time it receives a message containing an event (or it produces an event), it sends the event to its neighbours whose subscriptions match that event. Moreover, it gossips the message to other (non-interested) neighbours, so that the message is disseminated through the overlay even if none of its neighbours are subscribers for that event.

To analyze such protocol, we propose an analytical model which is based on complex network theory. 
The model estimates the amount of subscribers that may receive a given event. 
The approach is quite general; the network topology can be set by defining the node degree distribution probability. Depending on the network topology, the proportion of subscribers in the overlay, and the gossip probability threshold, it is possible to understand if the event reaches only a limited amount of nodes, or if it is spread through the whole network, i.e.~it might reach an infinite amount of nodes. 
Of course, this happens only when the network topology has a giant component.

In order to understand if the proposed mathematical model captures the main characteristics of the proposed system and validate its effectiveness, we have compared numerical outcomes with those obtained via simulation.
A discrete event simulator has been built, which is able to mimic the distributed communication protocol, on top of a randomly generated unstructured overlay whose topology can be specified using a given node degree probability distribution. We simulated a wide number of overlay networks, varying the network topology and degree distribution parameters, the size of the network, the portion of subscribers present in the system. We also varied the parameters characterizing the communication protocol, i.e.~the gossip probability. Results obtained via simulation are comparable with those coming from the analytical model.

The contribution of this paper can be summarized as follows.
\begin{enumerate}  
  \item We present a simple dissemination protocol that can be effectively employed over unstructured overlay networks to spread events. The protocol exploits local knowledge of peers about subscriptions made by their neighbour nodes, coupled with a gossip strategy. It can be employed quite effectively to build publish-subscribe systems on top of these easy-to-manage networks.
  \item We present an analytical model to characterize dissemination of events on top of unstructured networks. The overlay is modeled as a complex network. The model provides a general framework to understand if a generated event can percolate through the unstructured network.
  \item We employ the model to test the protocol over different overlay networks, and compare its results with those obtained via simulation. We focus on random networks built using a Poisson degree distribution, and on scale-free networks as well. We show that, also depending on the amount of subscribers, a small gossip probability is sufficient to spread events through the overlay. Hence, our outcomes demonstrate the viability of the use of unstructured networks to build large-scale, dynamic publish-subscribe systems. 
\end{enumerate}

In substance, the use of unstructured networks employing dissemination strategies based on local decision processes guarantees that the event percolates through the network. Thus, a node subscribing to a given type of contents will receive an event matching its subscriptions with high probability.
Of course, we are not suggesting here to replace completely structured and reliable distributed schemes, usually employed to build publish-subscribe services, with unstructured overlays using gossip. Rather, our claim is that this solution represents an interesting alternative when dealing with large scale and highly dynamic systems. In this case, in fact, the costs for managing and maintaining a structured (or centralized) distributed system is quite high. 

The remainder of this paper is organized as follows. Section \ref{sec:model} presents the system model. Section \ref{sec:protocol} states the local protocol executed at each node. Section \ref{sec:cn} presents the mathematical model. Section \ref{sec:exp} outlines results coming from a numerical analysis and simulation. Finally, Section \ref{sec:conc} provides some concluding remarks.

\section{System Model}\label{sec:model}

The system we consider is a \ac{P2P}~publish-subscribe system built on top of an unstructured overlay network. 
(Note that in the following we use the terms ``peer'' and ``node'' as synonyms.)
Peers are organized in a way that does not depend on the contents to be disseminated \cite{gridpeer}. 
Moreover, there is no central component that controls the dissemination of generated events. 

Each time a node produces a novel content to be published, it disseminates a message event containing it to its neighbours (the algorithm is explained in the next section). Each node receiving an event acts as a relay and forwards the event to other (neighbour) nodes. The dissemination is based on pure local decisions; in fact, peers employ a mixed strategy that combines gossip together with a local knowledge of subscriptions made their neighbours.

\subsection{Overlay Network}

We consider the set of nodes organized as a \ac{P2P}~overlay network. Each node $\mathbf{n}$ is connected to a given subset of nodes, 
whose number is specified using some probability distribution.\footnote{We use bold fonts to identify real entities in the distributed system, e.g.~host nodes or message events; all this in order to distinguish them from mathematical elements of the model, during the discussion.} We do not impose any restriction on the overlay, which can be generated using any kind of algorithm and attachment protocol executed when peers join the network. 
Hence, in general the overlay does not depend on the subscriptions made by peers in the~\ac{P2P} publish-subscribe system, i.e.~the overlay is unstructured.

We denote with $p_i$ the probability that a peer $\mathbf{n}$ has $i$ neighbours (the number of nodes connected to a node $\mathbf{n}$ is usually referred as its \emph{degree}). 
We assume that the overlay has a high number of nodes. This assumption comes from the fact that the solution we are studying is thought for very large and highly dynamical systems. If the number of nodes is low, or in presence of a relatively stable network, probably the use of an unstructured solution might be avoided, since other approaches can be proficiently employed, such as centralized solutions of structured distributed systems \cite{Ahullo:2008,Baldoni:2005,Eugster:2003}.
The high number of nodes, together with the random nature of contacts among peers in the overlay, augments the probability of having a low clustering in the network \cite{newmanHandbook}.

Events produced by publishers are included within messages spread through the overlay.
Direct communication may occur only between neighbour nodes. Hence, to disseminate information through the overlay, peers must act as relays and forward messages to their neighbours. 

It is clear that the topology of the overlay has a strong influence on the performance of the content dissemination \cite{disio11}. 
For instance, if a scale-free network is employed, then the network has a low diameter 
\cite{newman03thestructure}. 
However, a scale-free net contains a non-negligible fraction of peers, which maintain a high number of active connections, and hence they sustain a higher workload than the other low-degree nodes~\cite{Barabasi2000,guclu}. 
Conversely, if a network has uniform degree distribution, then the workload is equally shared among all peers. However, the diameter of the network increases, and so does the number of hops needed to
cover the whole network with a broadcast~\cite{gridpeer}. 
The framework employed in this work allows to assess how the topology of the overlay impacts the effectiveness of the distributed protocol by specifying the node degree probability distribution.
We focus on the network coverage and on the ability of the dissemination scheme to spread an event, depending on the topology of the overlay.

\subsection{P2P Publish-Subscribe System}

Peers in the overlay may act as \emph{subscribers} or \emph{publishers}. Subscribers register their interest in an event, or a pattern of events. Then, they must be notified asynchronously when events are generated by publishers \cite{Eugster:2003}. Such events may represent any kind of information which is usually filtered based on some event classification scheme. We are not going to describe in detail the plethora of existing methods to categorize events, since the particular approach is independent from the dissemination strategy, and hence not important for the purposes of this work. 
It is sufficient to assume that each event has some metadata associated to it, and that a subscription specifies a set of metadata the node is interested in (or predicates which allow to filter events).
Peers in the overlay may be subscribers and publishers at the same time, even for multiple patterns of events. If a peer in the overlay is not a subscriber nor a publisher for a given kind of content, it will act as a relay to disseminate these contents.

The protocol to disseminate events is completely decentralized. 
In our approach, each peer $\mathbf{n}$ stores in its cache all the subscriptions of its neighbours. Once $\mathbf{n}$ receives a message containing a given event $\mathbf{e}$, it is able to understand which neighbours are interested in receiving $\mathbf{e}$. Then, $\mathbf{n}$ sends $\mathbf{e}$ to all its neighbours that subscribed to that kind of event (if there are any). In addition, to avoid that $\mathbf{e}$ is discarded without having disseminated it, $\mathbf{n}$ gossips $\mathbf{e}$ to other remaining neighbours.
Nodes maintain in their caches information on messages which have been already handled, so as to avoid redundancy in the communication.

\section{The Protocol}\label{sec:protocol}

In the considered system, there are two main activities accomplished by peers. The first one is the subscription of a peer to a given type of events. The other activity is concerned with the publication and dissemination of a novel event. 

\subsection{The Subscription Protocol}

\begin{algorithm}[t]
\caption{Subscription protocol executed at node $\mathbf{n}$}
\label{alg:subscription}
\begin{small}
\begin{algorithmic}[1]
\STATE $N \leftarrow n$'s neighbours 
\STATE
\REQUIRE{Subscription for content type $\mathbf{c}$ from the application}\label{alg:sub_ff}
\STATE $msg = \langle ``subscription'', c \rangle$
\FORALL{$m \in N$} \hfill\COMMENT {send the subscription to all neighbours}
   \STATE \textsc{send}($msg, m$)%
\ENDFOR \label{alg:sub_ff_end}
\STATE
\REQUIRE{Subscription for content type $c$ removed from the application}\label{alg:sub_rr}
\STATE $msg = \langle ``remove'', c \rangle$
\FORALL{$m \in N$} \hfill\COMMENT {remove the subscription}
  \STATE \textsc{send}($msg, m$)%
\ENDFOR \label{alg:sub_rr_end}
\STATE
\REQUIRE{Reception of a control message from a peer $\mathbf{m}$}
\IF{subscription to any content type $c$}
\STATE \textsc{addInCache}($m, c$)\hfill\COMMENT {new subscription received}\label{alg:sub}
\ELSE \hfill\COMMENT {remove the subscription}\label{alg:sub_rr_sent}
\STATE \textsc{removeFromCache}($m, c$)\label{alg:rem}
\ENDIF
\end{algorithmic}
\end{small}
\end{algorithm}

The subscription protocol is very simple (see Algorithm \ref{alg:subscription}). When a peer $\mathbf{n}$ makes a novel subscription, it informs its neighbours (lines \ref{alg:sub_ff}--\ref{alg:sub_ff_end} in the algorithm). 

In turn, each node $\mathbf{m}$ receiving a message containing a novel subscription from a neighbour $\mathbf{n}$, adds a related entry in its neighbour table (line \ref{alg:sub} in the algorithm). This way, each time $\mathbf{m}$ receives an event $\mathbf{e}$ matching this subscription, $\mathbf{m}$ sends $\mathbf{e}$ to $\mathbf{n}$.

When a node is no more interested in a subscription, it informs its neighbours that will remove the related entry (lines \ref{alg:sub_rr}--\ref{alg:sub_rr_end} in the algorithm). In turn, upon reception at $\mathbf{n}$ of a control message from a node $\mathbf{m}$, stating that $\mathbf{m}$ is no more interested on a given subscription, that entry is removed from $\mathbf{n}$'s cache (line \ref{alg:rem}).

\subsection{The Dissemination Protocol}

The dissemination protocol
is a push scheme: nodes which have novel information to disseminate forward messages to other peers~\cite{disio11,disio10}. 
Algorithm \ref{alg:protocol} shows the pseudo-code of the algorithm executed at each peer $\mathbf{n}$ when an event $\mathbf{e}$ must be disseminated. The used notation is summarized in Table~\ref{tab:notation}.
It is worth mentioning that such code describes only the event management concerning the distribution of the event $\mathbf{e}$. We implicitly assume that another software module is in charge of analyzing the event $\mathbf{e}$, matching its metadata with the local subscriptions of the considered node, and in case passing the event to the application.

As to the event distribution service, once a given node $\mathbf{n}$ generates a novel event $\mathbf{e}$, or upon reception of a novel event $\mathbf{e}$ from a neighbour $\mathbf{m}$, $\mathbf{n}$ checks if it has already handled $\mathbf{e}$ in the past; in such a case, $\mathbf{n}$ drops $\mathbf{e}$ (lines \ref{alg:b_handled}--\ref{alg:e_handled}). This reduces the possibility that multiple copies of an event are processed and disseminated, thus limiting the amount of messages in the network.
The event is dropped also if the~\ac{TTL} associated to the event has reached a $0$ value. In this case, in fact, the event does not need to be forwarded elsewhere, since the maximum number of hops has been reached for that message. 

If $\mathbf{e}$ it is not dropped, $\mathbf{n}$ forwards it to the subset of neighbours 
whose subscriptions match the topics associated to $\mathbf{e}$, with exception of $\mathbf{m}$ (lines \ref{alg:b_f}--\ref{alg:e_f}).
Then, $\mathbf{n}$ considers the remaining set of its neighbours, i.e.~those nodes that are not interested in receiving $\mathbf{e}$.
For each node in this subset, $\mathbf{n}$ gossips $\mathbf{e}$ with a probability $\gamma \leq 1$ (lines \ref{alg:b_gossip}--\ref{alg:e_gossip}).

\begin{algorithm}[t]
\caption{Dissemination protocol executed at node $\mathbf{n}$}
\label{alg:protocol}
\begin{small}
\begin{algorithmic}[1]
\REQUIRE{Event $\mathbf{e}$ generated at $\mathbf{n}$ $\vee$ $\mathbf{e}$ received from a peer $\mathbf{m}$}
\STATE $e \leftarrow$ \textsc{removeFromBuffer}()
\IF{$\mathbf{e}$ already handled $\vee$ \textsc{TTL}($\mathbf{e}$) = $0$} \label{alg:b_handled} 
  \STATE Return
\ENDIF \label{alg:e_handled}
\STATE \textsc{decreaseTTL}($\mathbf{e}$)
\STATE $N \leftarrow n$'s neighbours $\setminus \ m$\hfill\COMMENT {$m =$ NULL if $\mathbf{e}$ originated at $\mathbf{n}$} \label{alg:b_f}
\STATE $I \leftarrow \{i | i \in N \wedge i\text{'s subscriptions match } e\}$
\FORALL{$i \in I$} \hfill\COMMENT {send $\mathbf{e}$ to all neighbour subscribers}
  \STATE \textsc{send}($e, i$)%
\ENDFOR \label{alg:e_f}
\FORALL{$i \in N \setminus I$} \hfill\COMMENT {gossip to the remaining neighbours}\label{alg:b_gossip}
     \IF{\textsc{random()} $< \gamma$}  %
         \STATE \textsc{send}($e,i$)%
     \ENDIF%
\ENDFOR \label{alg:e_gossip}
\end{algorithmic}
\end{small}
\end{algorithm}

\begin{table}[t]
\centering%
\begin{tabular}{rp{.8\columnwidth}}
\toprule
$\gamma :=$ & gossip probability\\
$f_i :=$ & probability that a node forwards an event to $i$ neighbours\\
$\overrightarrow{f}_i :=$ & probability that following a link, a node is reached that forwards an event to $i$ neighbours\\
$F :=$ & generating function of $f_i$\\
$\overrightarrow{F} :=$ & generating function of $\overrightarrow{f}_i$\\
$p_i :=$ & probability that a peer has degree equal to $i$\\
$q_i :=$ & excess degree probability, i.e.~probability that following a link a node is reached which has $i$ links other that the considered one\\
$\langle r \rangle :=$ & average number of nodes that receive an event\\
$r_i :=$ & probability that $i$ peers receive an event, starting from a given node\\
$\overrightarrow{r}_i :=$ & probability that $i$ peers receive an event, starting from a given link\\
$R :=$ & generating function of $r_i$\\
$\overrightarrow{R} :=$ & generating function of $\overrightarrow{r}_i$\\
$\sigma :=$ & probability of a subscription matching the considered event\\
$\langle s \rangle :=$ & average number of subscribers that receive an event\\
\bottomrule
\end{tabular}
\caption{Notation used in this paper}\label{tab:notation}
\end{table}

An important aspect is concerned with the~\ac{TTL} value, employed to avoid that messages are forwarded forever in the net. In particular, such~\ac{TTL} must be sufficiently large to guarantee that the message can be spread through the whole network.
An estimation of the network diameter (i.e.~the maximum number of hops required to reach a node starting from another one) can be obtained starting from the degree probability distribution, and in most kinds of nets it is usually a low number.
Hence, based on this common assumption, we will not consider such~\ac{TTL} value in the model presented in the next section. 

\section{Network Coverage}\label{sec:cn}

In this section, we analyze the performance of the decentralized~\ac{P2P} protocol presented in the previous section. We specifically focus on the coverage of the overlay, i.e.~the average amount of subscribers $\langle s \rangle$ that receive a given event $\mathbf{e}$. 
We denote with $\sigma$ the probability that a node has made a subscription matching $\mathbf{e}$, i.e.~$\sigma$ represents the portion of nodes in the overlay interested in receiving $\mathbf{e}$. 

We model each single event dissemination as a standalone activity. In other words, the model treats the distribution of generated events as independent tasks. This is a correct assumption if peers have a buffer whose size is sufficiently large to handle simultaneous events passing through it. Conversely, the model should be extended to consider possible buffer overflows. 

We consider networks with a large number of nodes. Following the approach presented in \cite{newman03thestructure,newmanHandbook}, we assume that links among nodes are randomly generated, based on a given node degree distribution \cite{BenderC78}. 
This does not represent a problem, since the overlays we are considering here are synthetic communication networks, which can be built using whatever algorithm chosen during the network design phase.
A consequence of the random nature of the attachment process is that, regardless of the node degree distribution, the probability that one of the second neighbours (i.e.~nodes at two hops from the considered node) is also a first neighbour of the same node, goes as $N^{-1}$, being $N$ the number of nodes in the overlay. Hence, this situation can be ignored since the number of nodes is high. 

\subsection{Degree and Excess Degree Distributions}

We denote with $p_i$ the probability that a peer $\mathbf{n}$ has degree equal to $i$. Starting from $\mathbf{n}$, another measure of interest is the number of connections (links) that a node $\mathbf{m}$, which is a neighbour of $\mathbf{n}$, may provide, other than the one that connects $\mathbf{m}$ with $\mathbf{n}$. In particular, the probability that, following a link in the overlay, we arrive to a peer $\mathbf{m}$ that has other $i$ links (hence its total degree is $i+1$) is
$$q_i = \frac{(i+1)p_{i+1}}{\sum_j j p_j}.$$
The probability $q_i$ is often referred as the \emph{excess degree distribution} \cite{newman03thestructure}. Probabilities $p_i$ and $q_i$ represent two similar concepts i.e.~the number of contacts of a considered peer (its degree), and the number of contacts obtained following a link (its excess degree), respectively. In the following, we introduce measures obtained by considering the degree $p_i$ of a node, and considering the excess degree $q_i$ of a link. In this last case, with a slight abuse of notation we denote all the probabilities/functions related to the excess degree with the same letter used for the degree, with an arrow on top of it, just to recall that the quantity refers to a link.

\subsection{Probability of Dissemination}

Given a peer $\mathbf{n}$ in charge of relaying an event $\mathbf{e}$, the probability that $\mathbf{n}$ forwards $\mathbf{e}$ to $i$ of its neighbours is
\begin{equation}\label{eq:f_k}
  f_i = [\sigma + (1-\sigma)\gamma]^i \sum_{j \geq i} p_j \binom{j}{i} [(1-\sigma)(1-\gamma)]^{j-i},
\end{equation}
which is obtained by considering all the possible cases of $\mathbf{n}$, having a degree higher than $i$, which forwards $\mathbf{e}$ to $i$ neighbours either because they are subscribers to events matching $\mathbf{e}$ (with probability $\sigma$), either because they are not subscribers but $\mathbf{n}$ decides to gossip $\mathbf{e}$ (with probability $(1-\sigma)\gamma$). Moreover, $\mathbf{n}$ does not gossip $\mathbf{e}$ to its remaining $j-i$ neighbours, which not subscribed to topics matching $\mathbf{e}$ (with probability $(1-\sigma)(1-\gamma)$).
In the rest of the discussion, for the sake of a more readable presentation, we denote $\Gamma = \sigma + (1-\sigma)\gamma$ and $1 - \Gamma = (1-\sigma)(1-\gamma)$.

A similar reasoning can be made to measure the probability that, following a link we arrive to a node that forwards $\mathbf{e}$ to $i$ other nodes. This probability is readily obtained by substituting, in (\ref{eq:f_k}) above, $p_j$ with $q_j$, i.e.
\begin{equation}\label{eq:g_k}
  \overrightarrow{f}_i = \Gamma^i \sum_{j \geq i} q_j \binom{j}{i} (1 - \Gamma)^{j-i}.
\end{equation}

To proceed with the reasoning, we need to introduce the generating functions for $f_i$, $\overrightarrow{f}_i$, as well as for $p_i$, $q_i$, i.e.
\begin{eqnarray}\label{eq:F_G} 
  G(x) = \sum_i p_i x^i, & & \overrightarrow{G}(x) = \sum_i q_i x^i, \\
  F(x) = \sum_i f_i x^i, & & \overrightarrow{F}(x) = \sum_i \overrightarrow{f_i} x^i.
\end{eqnarray}
In fact, if we consider the $F$ generating function, 
\begin{eqnarray}\label{eq:F_calcolo} 
  F(x) & = & \sum_i f_i x^i 
     =  \sum_i \Gamma^i x^i 
              \sum_{j \geq i} p_j \binom{j}{i} (1 - \Gamma)^{j-i}\nonumber \\
  & = & \sum_j p_j \sum_{i =0}^j \binom{j}{i}  \Gamma^i x^i (1 - \Gamma)^{j-i}\nonumber \\
   & = & \sum_j p_j (\Gamma x + 1 - \Gamma)^j\nonumber \\
  & = & G\big(\Gamma x + 1 - \Gamma\big)
\end{eqnarray}
One might notice that all the coefficients of the introduced generating functions are probabilities. In fact, $G(1) = \sum_i p_i = 1$, as well as $F(1) = \sum_i f_i = 1$, and so on.
Now, it is also possible to evaluate the average of the values $f_i$, by calculating the derivative of $f$ measured at $x=1$, since $F'(1) = \sum_i i f_i$ \cite{Wilf_1994}.
We have
\begin{eqnarray}\label{eq:F'(1)} 
  F'(x)\Bigl\lvert_{x=1} & = & \frac{dG}{dx}\big(\Gamma x + 1 - \Gamma\big)\Bigl\lvert_{x=1} 
                         =  \Gamma G'(1)\nonumber \\ 
                        & = & \Gamma \langle p \rangle,
\end{eqnarray}
where $\langle p \rangle$ is the mean node degree probability.

From a similar reasoning,
\begin{eqnarray}\label{eq:F_over'(1)} 
  \overrightarrow{F}'(x)\Bigl\lvert_{x=1}
  = \Gamma \overrightarrow{G}'(1) = \Gamma \langle q \rangle,
\end{eqnarray}
where $\langle q \rangle$ is the mean value of the excess degree, that is \cite{newmanHandbook}
\begin{eqnarray}\label{eq:mean_q} 
  \langle q \rangle & = & \sum_i i q_i = \frac{\sum_i i (i+1) p_{i+1}}{\sum_j j p_j}
   =  \frac{\sum_i (i-1) i p_i}{\sum_j j p_j} \nonumber \\
  & = & \frac{\langle p^2 \rangle - \langle p \rangle}{\langle p \rangle}.
\end{eqnarray}

\subsection{Number of Receivers and Subscribers}

We can now consider the whole number of nodes reached by a message starting from a given node, regardless of the number of hops. Let denote with $r_i$ the probability that $i$ peers receive an event, starting from a given node. Similarly, denote with $\overrightarrow{r}_i$ the probability that $i$ peers are reached by the event dissemination, starting from a link. In general, $\overrightarrow{r}_i$ can be defined using the following recurrence,
\begin{align}\label{eq:r_k_l}
  \overrightarrow{r}_0 & =  0,\nonumber \\
  \overrightarrow{r}_{i+1} & =  \sum_{j \geq 0} \overrightarrow{f}_j \sum_{a_1 + a_2 + \ldots + a_j = i} \overrightarrow{r}_{a_1} \overrightarrow{r}_{a_2} \ldots \overrightarrow{r}_{a_j}.
\end{align}
Equation (\ref{eq:r_k_l}) can be explained as follows. It measures the probability that following a link we disseminate the event to $i+1$ peers. (The case $\overrightarrow{r}_0$ is impossible, since at the end of a link there must be a node.) In general, one peer is the one reached at the end of the link itself. Then, we consider the probability that the peer has other $j$ links (varying the value of $j$). Each link $k$ allows to disseminate the event to $a_k$ peers, and the sum of all these reached peers equals to $i$.

Similarly, we can calculate $r_k$ as follows
\begin{align}\label{eq:r_k}
  r_0  &= 0,\nonumber \\
  r_{i+1}  &= \sum_{j \geq 0} f_j \sum_{a_1 + a_2 + \ldots + a_j = i} \overrightarrow{r}_{a_1} \overrightarrow{r}_{a_2} \ldots \overrightarrow{r}_{a_j}.
\end{align}
In this case, we start from the peer itself, considering it has a degree equal to $j$; and as before, from its $j$ links we can reach $i$ other peers, globally.

The use of generating functions may be of help to handle these two equations \cite{Wilf_1994}. In fact, if we consider the generating functions for $r_i$ and $\overrightarrow{r}_i$,
\begin{eqnarray}\label{eq:R} 
  R(x) = \sum_i r_i x^i, & & \overrightarrow{R}(x) = \sum_i \overrightarrow{r}_i x^i
\end{eqnarray}
then, after some manipulation typical for generating functions (e.g.~\cite{newmanHandbook}) we arrive to the following result
\begin{eqnarray}\label{eq:R_q_rec} 
  \overrightarrow{R}(x)  =  x \sum_{j \geq 0} \overrightarrow{f}_j [\overrightarrow{R}(x)]^j
     =  x \overrightarrow{F}(\overrightarrow{R}(x))
\end{eqnarray}
and, similarly,
\begin{eqnarray}\label{eq:R_rec} 
  R(x)  =  x \sum_{j \geq 0} f_j [\overrightarrow{R}(x)]^j
     =  x F(\overrightarrow{R}(x)).
\end{eqnarray}
From the generating functions, we might recover the elements $r_i$, $\overrightarrow{r}_i$ composing them. Unfortunately, equations (\ref{eq:R_q_rec}), (\ref{eq:R_rec}) may be difficult to solve, depending on the degree probability distribution $p_i$ which controls the whole introduced measures \cite{newmanHandbook}.

But actually, we are not interested that much in the single values of $r_i$, $\overrightarrow{r}_i$. In fact, it is easier and more useful to measure the average number $\langle r \rangle$ of peers that receive a given event through the dissemination protocol. To this aim, we can employ the typical formula for generating functions
$\langle r \rangle = R'(1)$ \cite{Wilf_1994}.
In fact, taking the first equation of (\ref{eq:R}), differentiating and evaluating the result for $x=1$, and since $r_0 = 0$, we have 
$$R'(x) \Bigl\lvert_{x=1} = \sum_i i r_i,$$
which is the mean value related to the distribution of $r_i$ coefficients.
We already observed that the coefficients of the introduced generating functions are probabilities, and thus
$F(1)= \sum_i f_i = 1$, and similarly $\overrightarrow{F}(1)=1$, $R(1)=1$, $\overrightarrow{R}(1)=1$. Hence, taking (\ref{eq:R_rec}) and differentiating 
\begin{eqnarray}\label{eq:r} 
\langle r \rangle & =  & R'(1) = \big[F(\overrightarrow{R}(x)) + x F'(\overrightarrow{R}(x))\overrightarrow{R}'(x)\big]_{x=1}\nonumber \\
 &  = & 1 + F'(1)\overrightarrow{R}'(1).
\end{eqnarray}
Similarly, from (\ref{eq:R_q_rec}),
\begin{eqnarray}\label{eq:R_q_der} 
\overrightarrow{R}'(1) & = & \big[\overrightarrow{F}(\overrightarrow{R}(x)) + x \overrightarrow{F}'(\overrightarrow{R}(x))\overrightarrow{R}'(x)\big]_{x=1}\nonumber \\
 & = & 1 + \overrightarrow{F}'(1)\overrightarrow{R}'(1).
\end{eqnarray}
Thus,
\begin{equation}\label{eq:R_q_der_f}
\overrightarrow{R}'(1) = \frac{1}{1 - \overrightarrow{F}'(1)}.
\end{equation}
This last equation allows to find the final formula for $\langle r \rangle$,
\begin{eqnarray}\label{eq:r_final}
\langle r \rangle & = & 1 + \frac{F'(1)}{1 - \overrightarrow{F}'(1)}\nonumber \\ 
 &= & 1 + \frac{\Gamma\langle p \rangle^2}{(1+\Gamma)\langle p \rangle - \Gamma \langle p^2 \rangle}. 
\end{eqnarray}

Now, $\langle r \rangle$ is the number of peers that receive the event, regardless if they are subscribers or simply relay nodes. To obtain the average number of subscribers $\langle s \rangle$ that receive the event, it suffices to multiply $\langle r \rangle$ by the probability that a peer is a subscriber $\sigma$, hence obtaining
$$\langle s \rangle = \sigma \langle r \rangle.$$

\subsection{Percolation Probability}

As it is quite typical in complex network theory, it is actually easier to examine infinite networks rather than just large ones. The analysis of infinite networks, under conditions similar to those of large scale networks, allows to understand important peculiarities of the real networks and on protocols executed by their nodes. For instance, it is possible to understand if a message can percolate through the network.
This assumption is perfectly reasonable in our scenario, since we consider very large dynamical systems (with a number of nodes that tends to infinity) where peers know only their neighbours and manage contents based on local knowledge about nodes' subscriptions.

Equation (\ref{eq:r_final}) has a divergence when $(1+\Gamma)\langle p \rangle = \Gamma \langle p^2 \rangle$, which signifies that the event reaches an infinite number of nodes, i.e.~the event percolates through the network.
By looking at the parameters, this situation depends, first, on the nodes' connectivity, i.e.~the node degree probability distribution $p_i$. In fact, the degree probability distribution determines if the overlay has a giant component (i.e.~the largest subset of connected nodes which scales with the network size, and thus has a number of nodes whose limit tends to $\infty$), rather than being partitioned into a set of components of limited size \cite{newmanHandbook}. 
The event can be spread to a large (infinite) number of nodes only when there is such a giant component; otherwise, i.e.~when the network is partitioned into a high number of components of limited size,
the event can be sent to a limited number of nodes only. Studies exist that allow to understand how to build networks with a giant component \cite{Flaxman:2005,newmanHandbook}.

Second, the value of $\sigma$ has influence on both the number of subscribers to be reached and on the dissemination of events. In fact, the higher $\sigma$ the higher the probability that a node has some of its neighbours which are subscribers to a given type of events; these nodes will be receivers of the event and subsequently they will act as relays for such event. 

Third and final, the gossip probability $\gamma$ determines if the message event is spread through the network even when the amount of subscribers in the overlay for a given event type is small, i.e.~when $\sigma$ has a very low value. Of course, setting $\gamma = 1$ allows to flood the event to the whole component (from which the event has been originated). This is a fair choice when the network has a tree-like structure, or when the network has a very low clustering. Conversely, a low value for $\gamma$ should be employed when there are loops in the overlay.

A completely different scenario is concerned with the situation when the network is formed by limited clusters only (there is no giant component). In such a case, in fact, the number of reached nodes does not grow proportionally with the network size, and a finite number of subscribers might receive a published event.

\section{Experimental Results}\label{sec:exp}

This section presents an assessment performed to validate the model discussed in the previous section and evaluate the ability of the outlined \ac{P2P}~publish-subscribe system to disseminate contents. The evaluation is performed by considering the analytical model and results obtained through a simulation of the distributed protocol. The two approaches provide similar outcomes. In particular, when the theoretical model estimates that an infinite amount of nodes is reached through the dissemination, simulations show that a significant portion of the simulated network receives the events, as expected.

The focus here is on network coverage. Another important metric to consider is the number of sent messages. In this sense, the protocol ensures that peers disseminate a given event at most once.
Moreover, the tree-like structure of the overlay limits that multiple copies of the same event are received by a peer.

\begin{figure}[t]
   \centering
   \includegraphics[angle=-90,width=\linewidth]{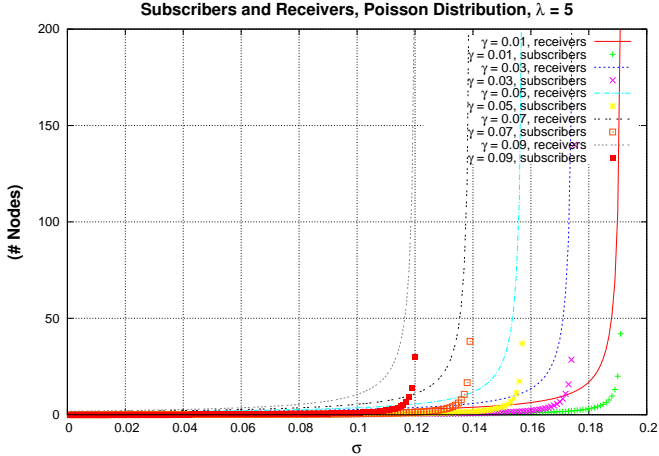}
   \caption{Number of receivers and subscribers: topology based on a Poisson degree distribution with mean $\lambda=5$.}
   \label{fig:poisson_l5_sub}
\end{figure}

\begin{figure}[t]
   \centering
   \includegraphics[angle=-90,width=\linewidth]{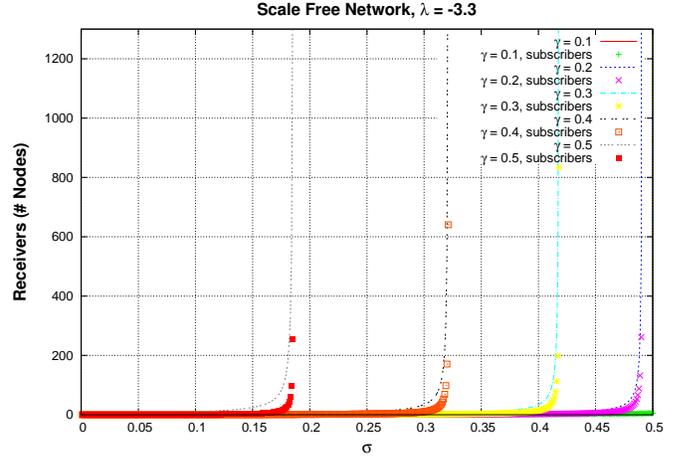}
   \caption{Number of receivers and subscribers: scale-free topology with exponent $\lambda=-3.3$.}
   \label{fig:sf_sub}
\end{figure}

\subsection{Theoretical Model}

We employed the framework presented in Section \ref{sec:cn} to assess the performance of the dissemination protocol, based on the overlay network topology, i.e.~node degree distribution, the subscription probability $\sigma$ and the gossip probability $\gamma$. 
Figure \ref{fig:poisson_l5_sub} shows the number of nodes receiving an event, spread through the network, when the unstructured overlay has a topology based on a Poisson node degree distribution with mean value $\lambda=5$ (we tested the framework with other $\lambda$ values, obtaining similar results). 
Lines in the chart correspond to the whole number of receivers (i.e.~relay nodes and subscribers), while points correspond to the number of subscribers.
Results are obtained varying the value of $\sigma$ (on the x-axis), i.e.~the portion of subscribers present in the overlay.

From these two figures it is easy to see that, for each specific $\gamma$ value, there is a phase transition, i.e.~as $\sigma$ is varied there is an abrupt increment on the number of receivers (and subscribers), passing from a limited value to $\infty$, i.e.~the event percolates through the network. 
This phase transition depends on the parameters used to set the distributed system. In fact, the value of $\sigma$ not only represents the subscription probability, but it influences also the event dissemination in the overlay (a node forwards with probability $1$ the event to each of its neighbours that subscribed to that event).
Finally, the value of $\gamma$ does not change the trend of the curves; basically, the higher $\gamma$ the smaller the value of $\sigma$ to have a transition. 

Similar considerations can be made for Figure \ref{fig:sf_sub}, where the estimated amount of receivers and subscribers is reported for a scale-free network with a degree distribution $\sim p^\lambda$, with $\lambda = -3.3$. Also in this case, each curve corresponds to a specific $\gamma$ value, while varying $\sigma$. The chart shows that for each curve there is a phase transition, where the number of receiving nodes passes from a limited (low) value to an infinite number.

\begin{figure}[t]
   \centering
   \includegraphics[angle=-90,width=\linewidth]{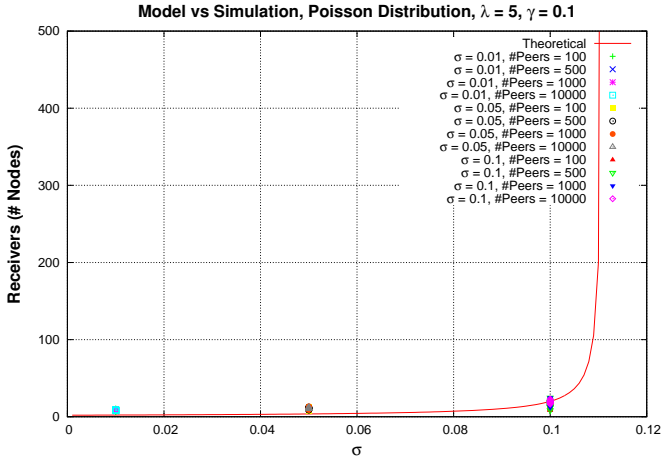}
   \caption{Model vs Simulation: topology based on a Poisson degree distribution with mean $\lambda = 5$.}
   \label{fig:confronto_g1}
\end{figure}

\begin{figure}[t]
   \centering
   \includegraphics[angle=-90,width=\linewidth]{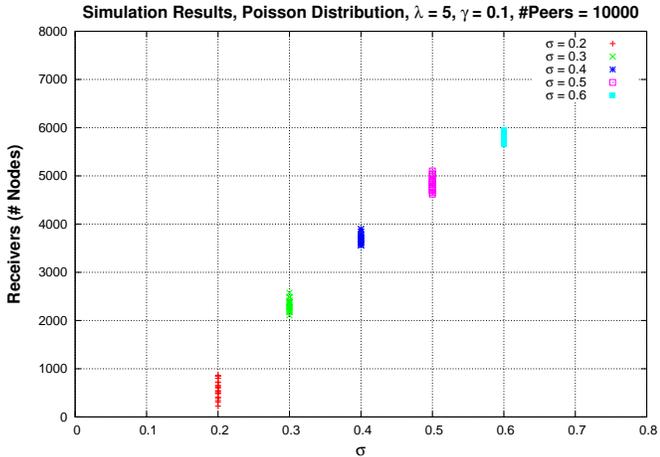}
   \caption{Model vs Simulation: Poisson degree distribution, $\lambda = 5$, varying $\sigma$ above the phase transition. The chart reports the number of receiving nodes through simulation. The theoretical model returns an infinite number of nodes (being the mo\-del\-led overlay an infinite graph), not shown here.}
   \label{fig:confronto_s_above}
\end{figure}

\begin{figure}[t]
   \centering
   \includegraphics[angle=-90,width=\linewidth]{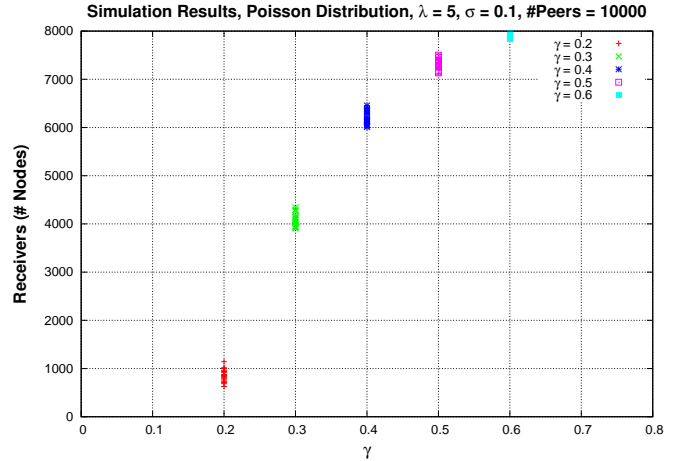}
   \caption{Model vs Simulation: Poisson degree distribution, $\lambda = 5$, varying $\gamma$ above the phase transition. Number of receiving nodes obtained through simulation (the model returns an infinite sub-graph).}
   \label{fig:confronto_g_above}
\end{figure}

\subsection{Simulation}

In order to assess the theoretical model proposed in the paper, we have built a discrete-event simulator mimicking the presented protocol. The simulator was written in C code. Pseudo-random number generation was performed by employing the GNU Scientific Library,  a library that provides implementation of several mathematical routines for numerical and statistical analysis \cite{gsl-web-2010}. The simulator allows to test the behavior of a given amount of nodes executing a publish-subscribe distributed system employing the protocol explained in Section \ref{sec:protocol}.

The simulator allows to generate a random network based on the chosen degree distribution. In particular, once having assigned a specific target degree to each node, using the selected degree distribution, a random mapping is made so that links are created until each node has reached its own target degree.
The simulator was set to manage the dissemination of a single type of events.
During the initialization phase, for each node a random choice was made, in order to set that node as a subscriber of the event type or not, based on the probability $\sigma$.

We varied the network topology, the number of nodes and statistical parameters characterizing the network degree distribution. For each network setting, we repeated the simulation using a corpus of $20$ different randomly generated networks. For each network, we analyzed the dissemination of $400$ events published by random nodes. In the results that follow, for each generated network we show the average number of receiving nodes, i.e.~subscribers and relays; this number allows to understand if the distributed protocol is able to disseminate the event through the unstructured network, using the presented protocol. 

\subsubsection{Poisson Degree Distribution}

Here, we show results for networks generated through a Poisson degree distribution.
Figure \ref{fig:confronto_g1} shows results obtained from simulation and the theoretical model. We simulated different corpuses of networks, varying the number of nodes and the value of the gossip probability $\gamma$. 
Each point in the chart corresponds to the average number of receivers for a simulated network. The line corresponds to the theoretical value measured using equation (\ref{eq:r_final}). It is possible to observe that all results from the simulations lye near the theoretical value, regardless on the considered number of simulated network nodes. Hence, the model is able to capture the behavior of the distributed protocol.

Figures \ref{fig:confronto_s_above}, \ref{fig:confronto_g_above} show results obtained in our simulations when $\gamma=0.1$ (resp.~$\sigma=0.1$), while varying $\sigma$ (resp.~$\gamma$), above the phase transition. According to the model, the system is above the phase transition. Hence, assuming an infinite number of nodes in the network, an infinite number of receivers is reached. As concerns simulations, instead, we expect that a non-negligible portion of nodes is reached during the dissemination of an event. Of course, since the dissemination is based on rather low values of $\gamma, \sigma$ probabilities, and since the network clustering of these considered networks is quite low (we employ a random attachment process to build links in the network \cite{newman03thestructure,newmanHandbook}),
it is unlikely that all network nodes receive the event being disseminated.
In fact, because of the tree-like structure of the network, every time we decide not to exploit a link, we might cut away some branch (and consequently some sub-graph) of the overlay.
Indeed, results confirm our outlook. 
A non-negligible portion of nodes is reached in each configuration. Yet, the whole overlay is not covered completely. The amount of the reached nodes increases with the varied parameter $\sigma$ (resp.~$\gamma$). Of course, the entire network (or at least, the component to which the node belongs) can be reached by flooding the event.

Similar results were obtained for different networks built varying the statistical parameters of the random graph. In substance, all this means that the protocol is able to spread a given event in the network in random graphs with Poisson degree distributions.

\subsubsection{Scale-Free Networks}

Scale free networks gained a lot of interest in recent years. These networks are characterized by a degree distribution following a power law. 
They are characterized by the presence of hubs, i.e.~nodes with degrees higher than the average, that have an important impact on the connectivity of the net.
The interest on scale-free networks in this work relates to the fact that several peer-to-peer systems are indeed scale-free networks \cite{simutools,newman03thestructure}.

\begin{figure}
   \centering
   \includegraphics[angle=270,width=\linewidth]{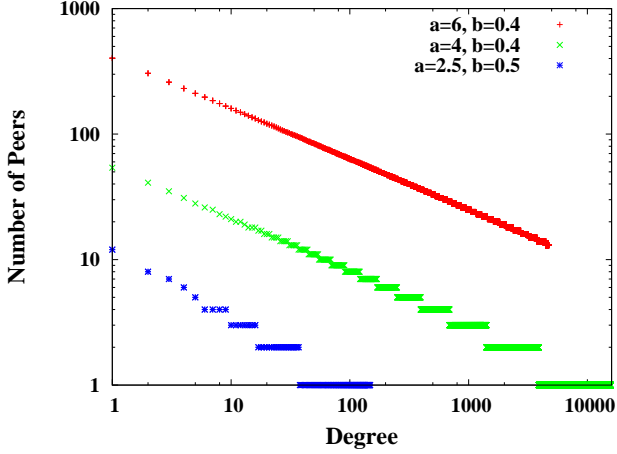}
   \caption{Degree Distribution of some scale-free networks using the construction method proposed in \cite{Aiello00arandom}}
   \label{fig:fig_rete_Aiello}
\end{figure}

\begin{figure}[t]
   \centering
   \includegraphics[angle=-90,width=\linewidth]{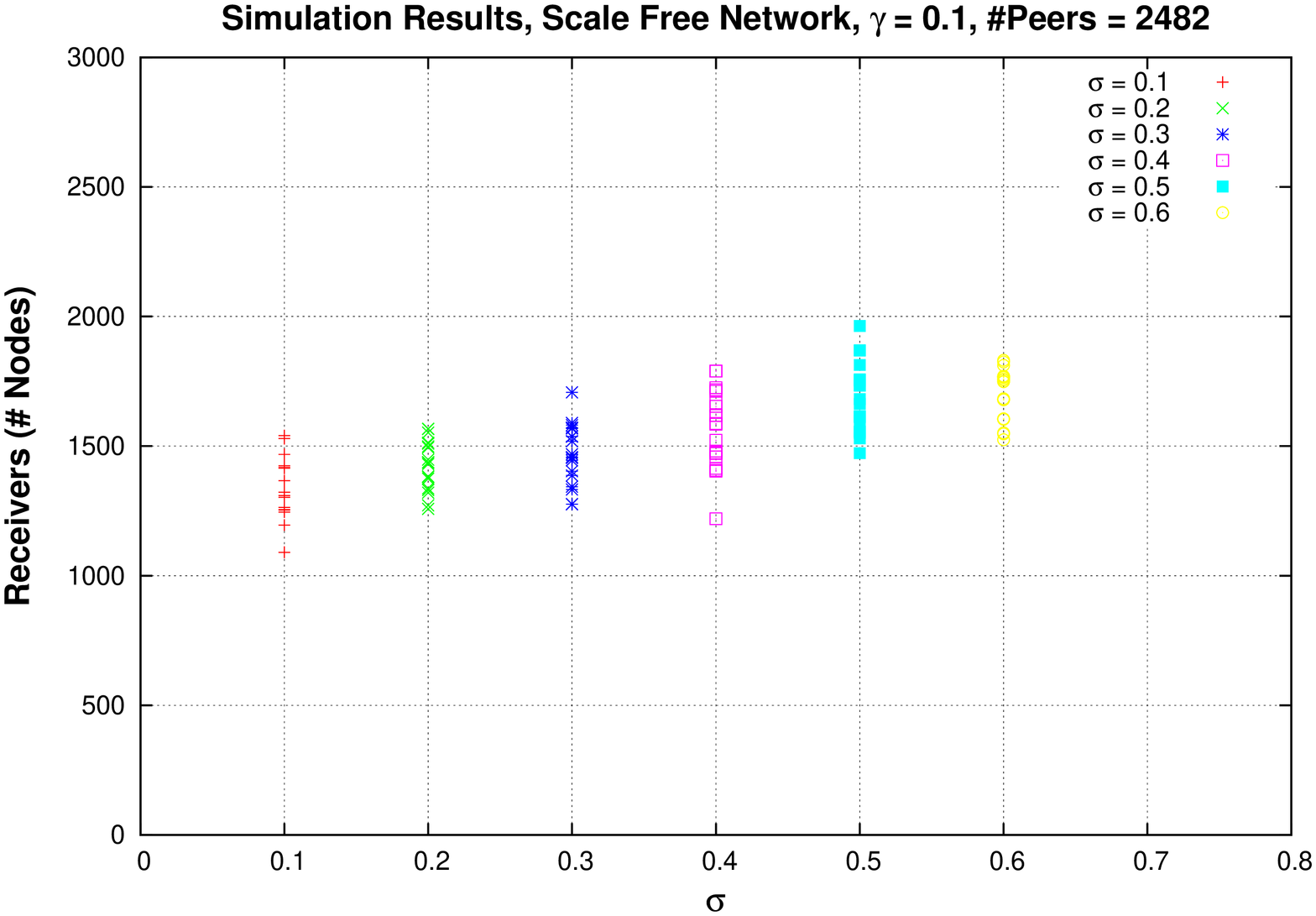}
   \caption{Model vs Simulation: scale free network, $a = 6, b = 1$, varying $\sigma$ above the phase transition. Number of receiving nodes obtained through simulation (the model returns an infinite sub-graph).}
   \label{fig:SF_confronto_g_above}
\end{figure}

\begin{figure}[t]
   \centering
   \includegraphics[angle=-90,width=\linewidth]{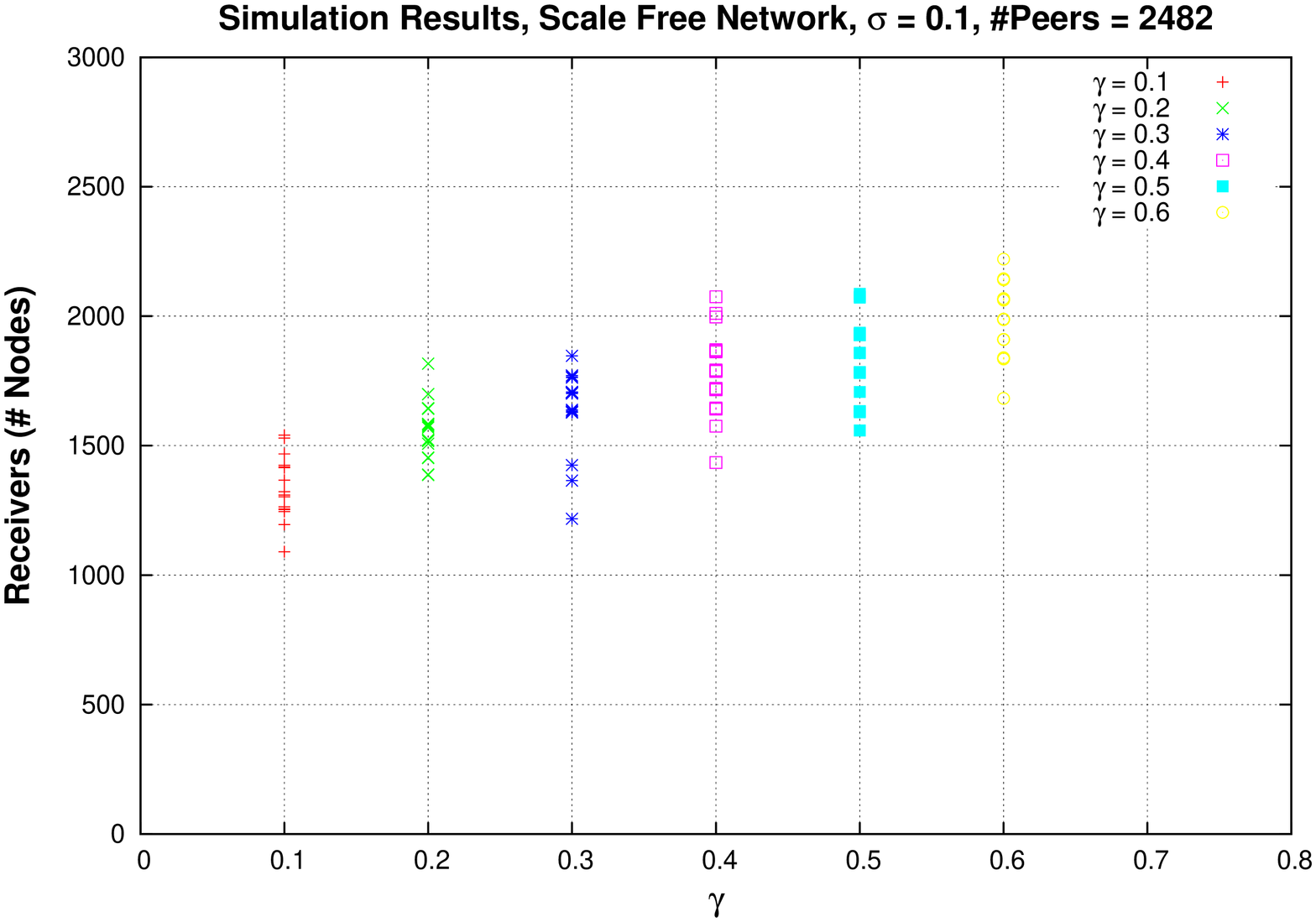}
   \caption{Model vs Simulation: scale free network, $a = 6, b = 1$, varying $\gamma$ above the phase transition. Number of receiving nodes obtained through simulation (the model returns an infinite sub-graph).}
   \label{fig:SF_confronto_s_above}
\end{figure}

\begin{figure}[t]
   \centering
   \includegraphics[angle=-90,width=\linewidth]{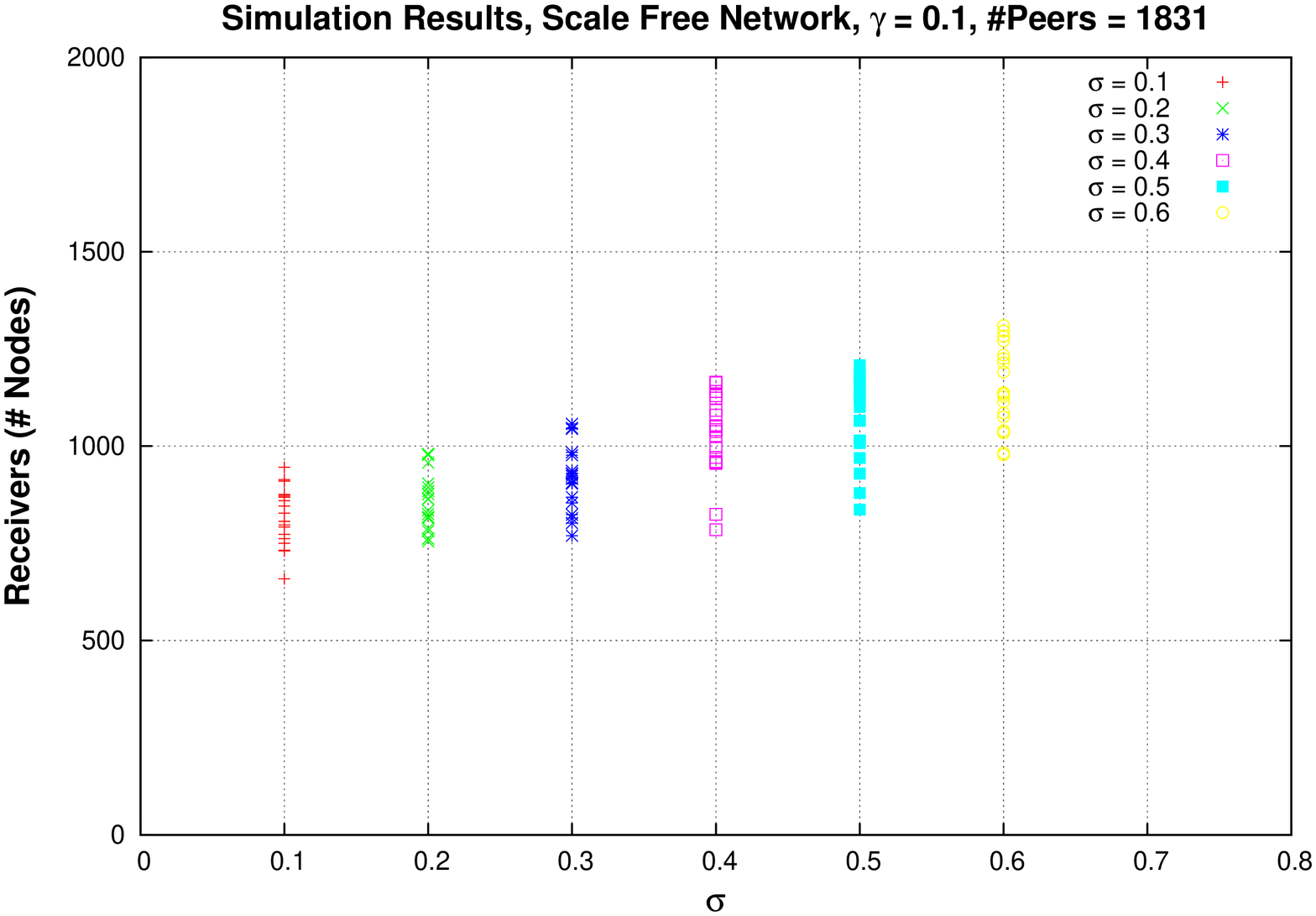}
   \caption{Model vs Simulation: scale free network, $a = 6, b = 1.1$, varying $\sigma$ above the phase transition. Number of receiving nodes obtained through simulation (the model returns an infinite sub-graph).}
   \label{fig:SF_confronto_g_beta1.1}
\end{figure}

\begin{figure}[t]
   \centering
   \includegraphics[angle=-90,width=\linewidth]{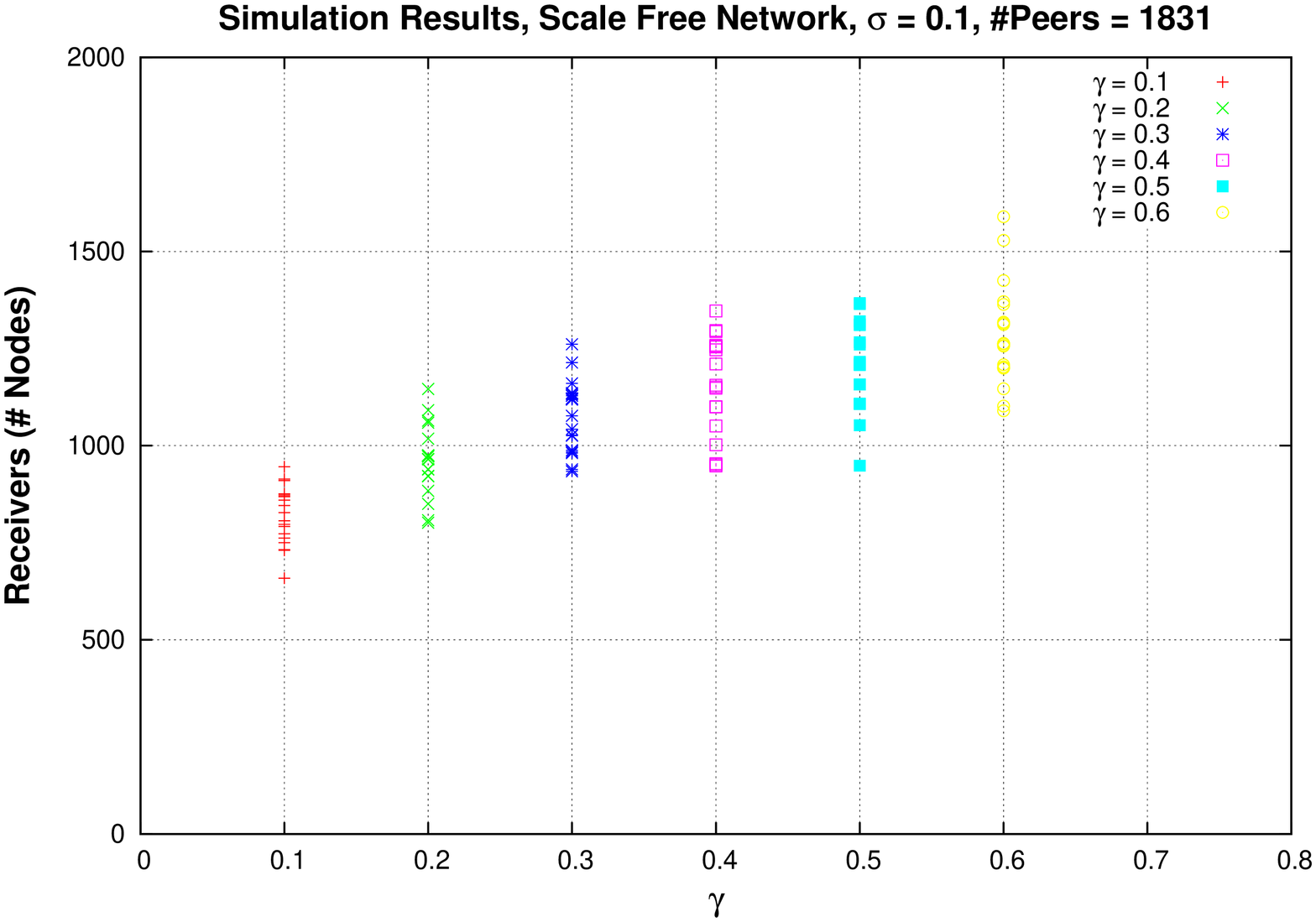}
   \caption{Model vs Simulation: scale free network, $a = 6, b = 1.1$, varying $\gamma$ above the phase transition. Number of receiving nodes obtained through simulation (the model returns an infinite sub-graph).}
   \label{fig:SF_confronto_s_beta1.1}
\end{figure}

\begin{figure}[t]
   \centering
   \includegraphics[angle=-90,width=\linewidth]{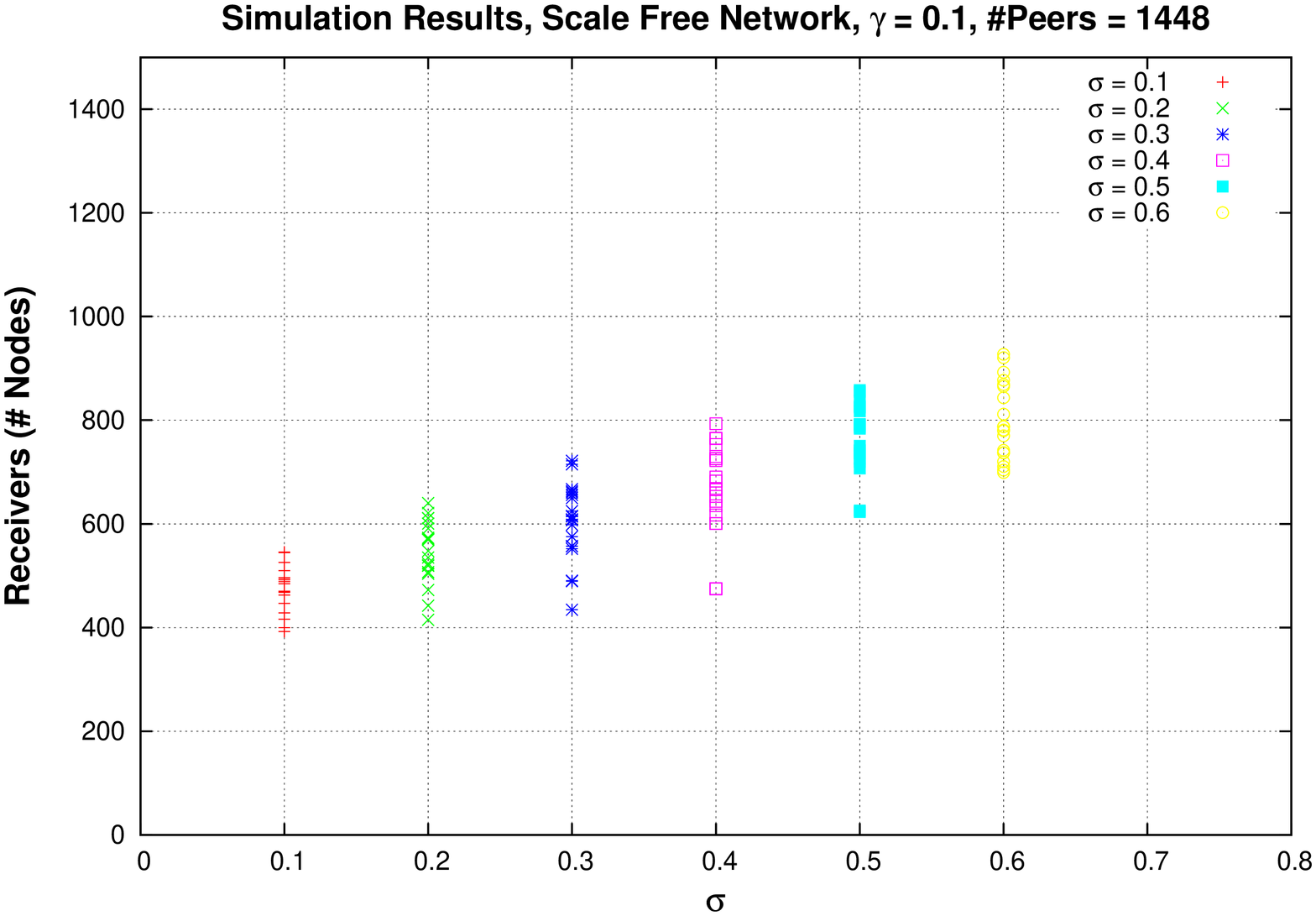}
   \caption{Model vs Simulation: scale free network, $a = 6, b = 1.2$, varying $\sigma$ above the phase transition. Number of receiving nodes obtained through simulation (the model returns an infinite sub-graph).}
   \label{fig:SF_confronto_g_beta1.2}
\end{figure}

\begin{figure}[t]
   \centering
   \includegraphics[angle=-90,width=\linewidth]{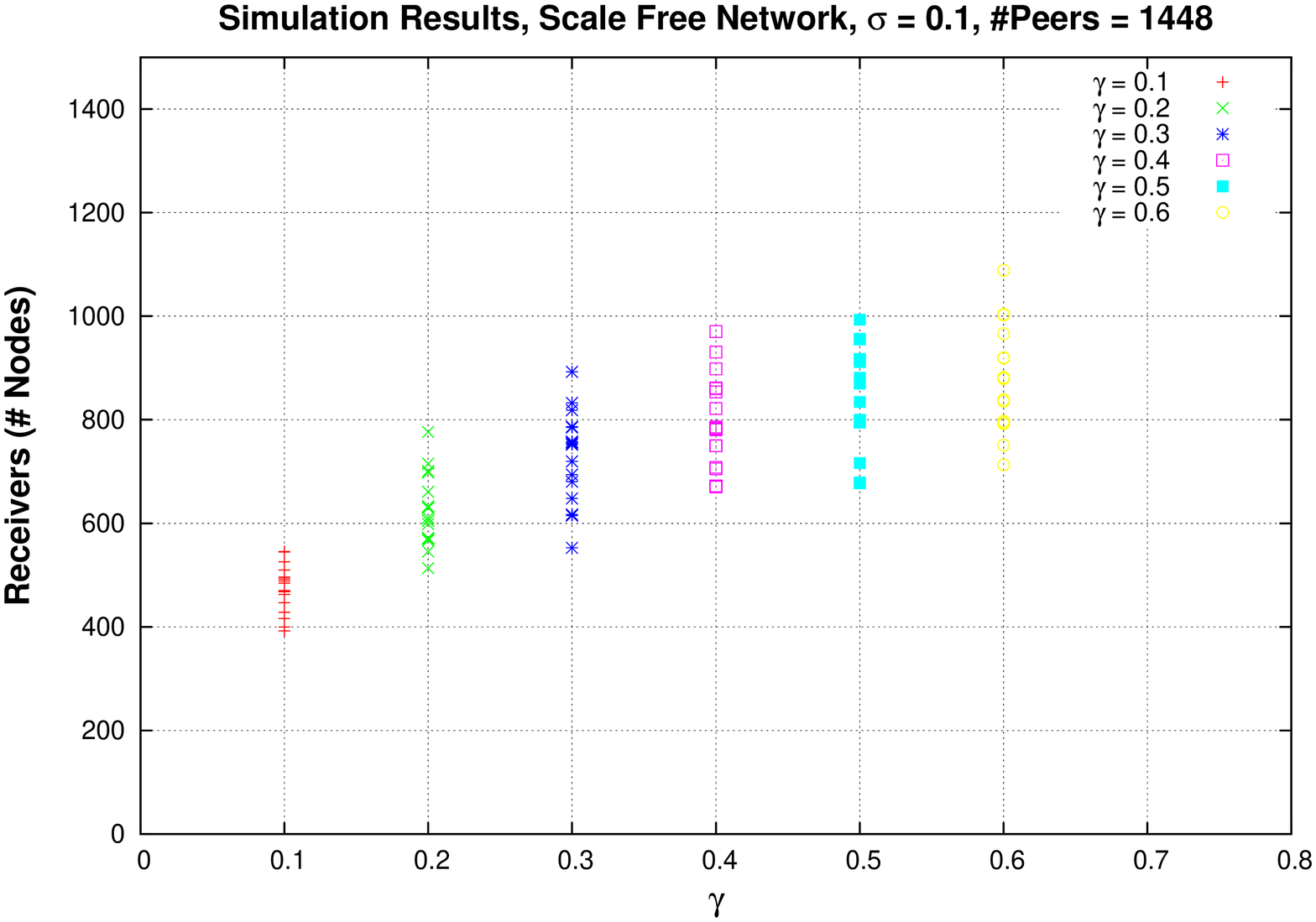}
   \caption{Model vs Simulation: scale free network, $a = 6, b = 1.2$, varying $\gamma$ above the phase transition. Number of receiving nodes obtained through simulation (the model returns an infinite sub-graph).}
   \label{fig:SF_confronto_s_beta1.2}
\end{figure}

\begin{figure}[t]
   \centering
   \includegraphics[angle=-90,width=\linewidth]{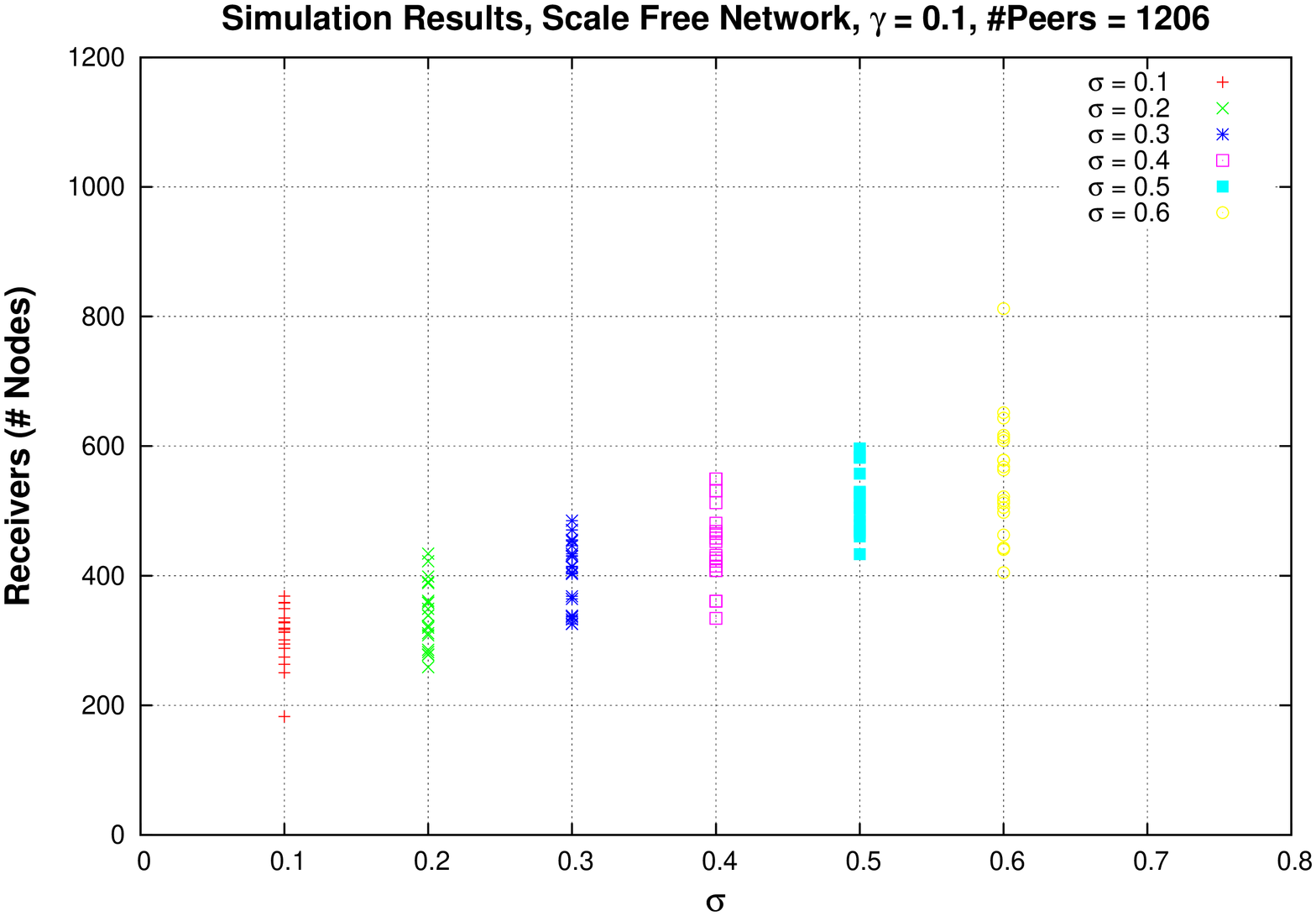}
   \caption{Model vs Simulation: scale free network, $a = 6, b = 1.3$, varying $\sigma$ above the phase transition. Number of receiving nodes obtained through simulation (the model returns an infinite sub-graph).}
   \label{fig:SF_confronto_g_beta1.3}
\end{figure}

\begin{figure}[t]
   \centering
   \includegraphics[angle=-90,width=\linewidth]{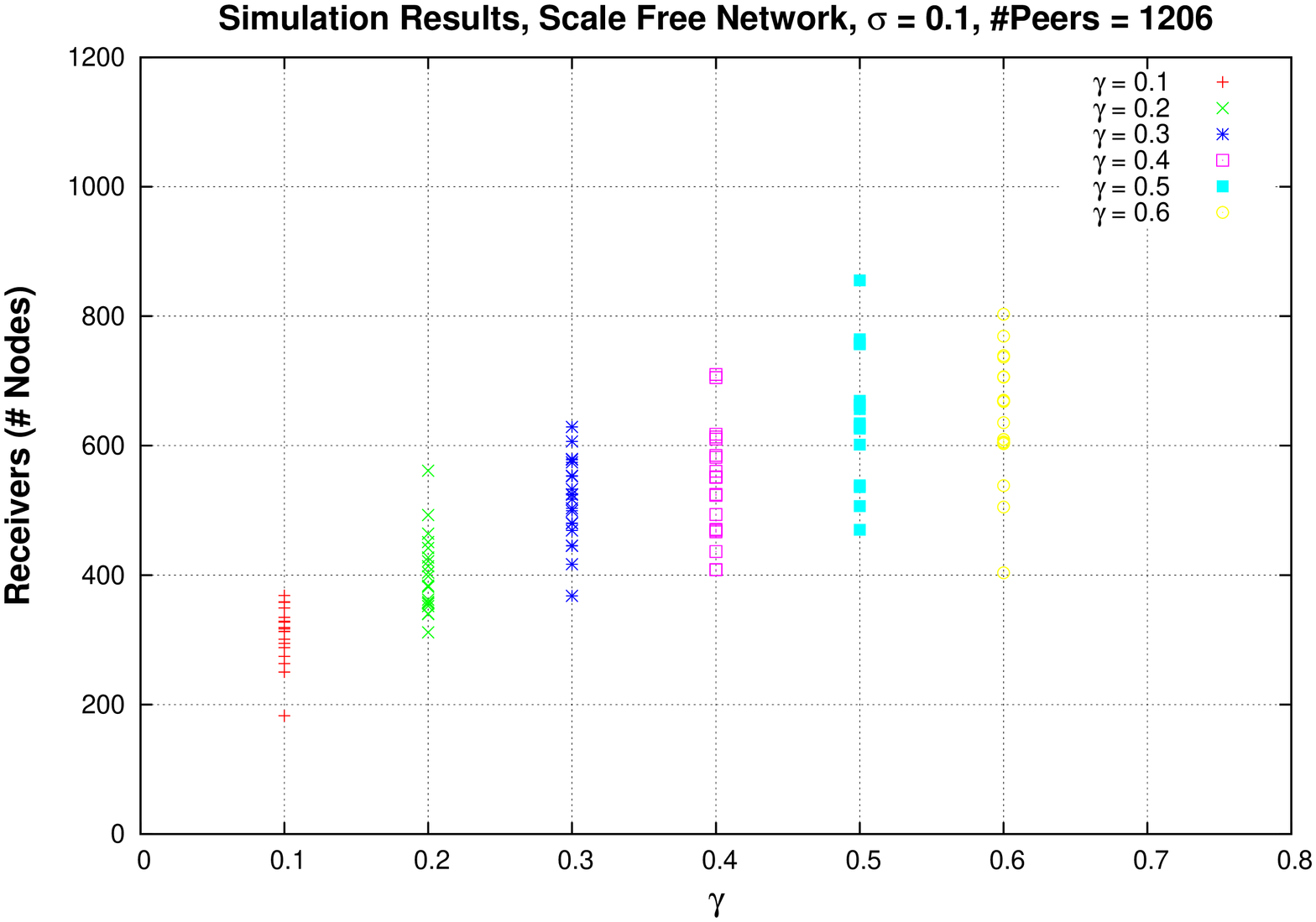}
   \caption{Model vs Simulation results: scale free network, $a = 6, b = 1.3$, varying $\gamma$ above the phase transition. Number of receiving nodes obtained through simulation (the model returns an infinite sub-graph).}
   \label{fig:SF_confronto_s_beta1.3}
\end{figure}

\begin{figure}[t]
   \centering
   \includegraphics[angle=-90,width=\linewidth]{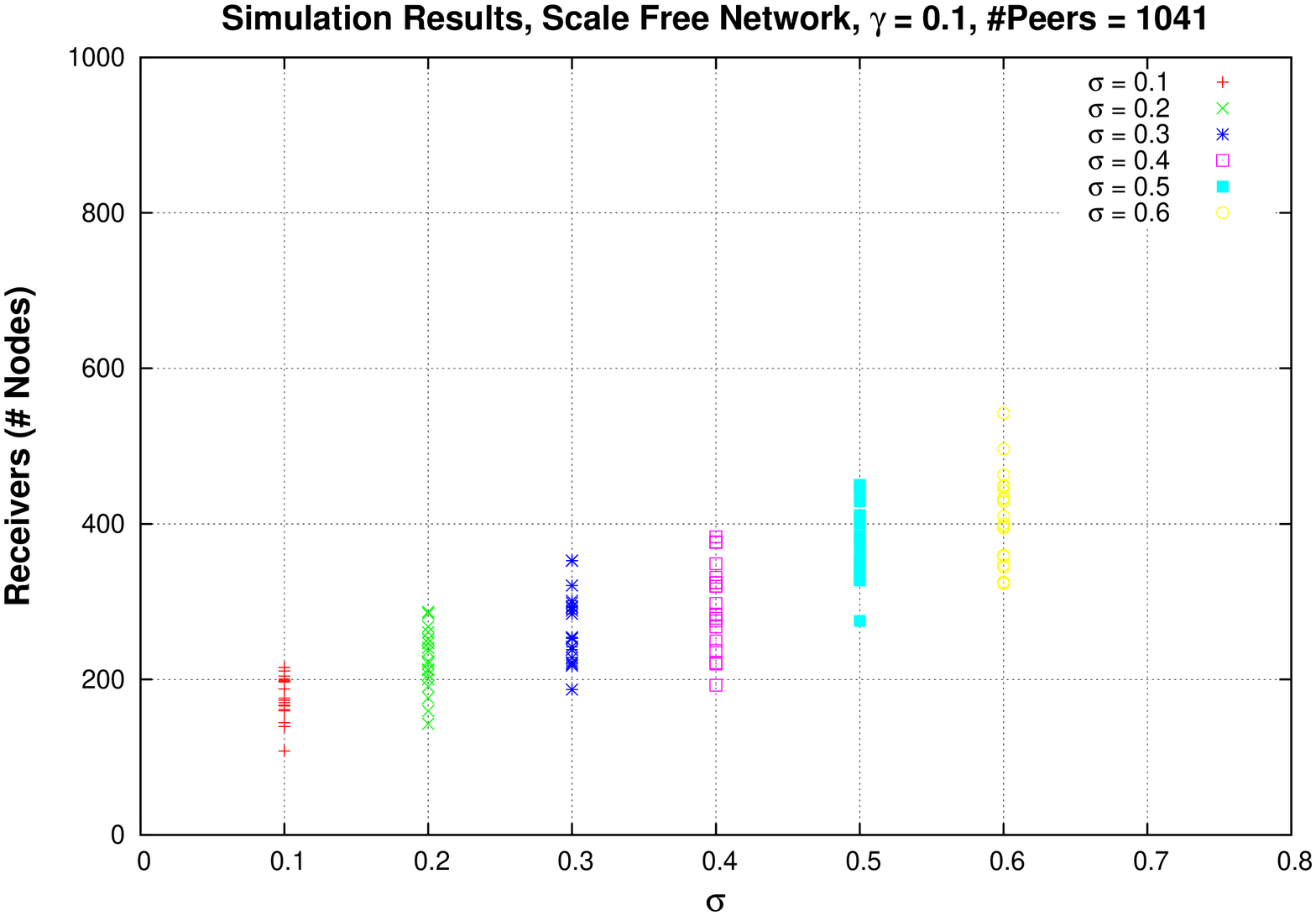}
   \caption{Model vs Simulation: scale free network, $a = 6, b = 1.4$, varying $\sigma$ above the phase transition. Number of receiving nodes obtained through simulation (the model returns an infinite sub-graph).}
   \label{fig:SF_confronto_g_beta1.4}
\end{figure}

\begin{figure}[t]
   \centering
   \includegraphics[angle=-90,width=\linewidth]{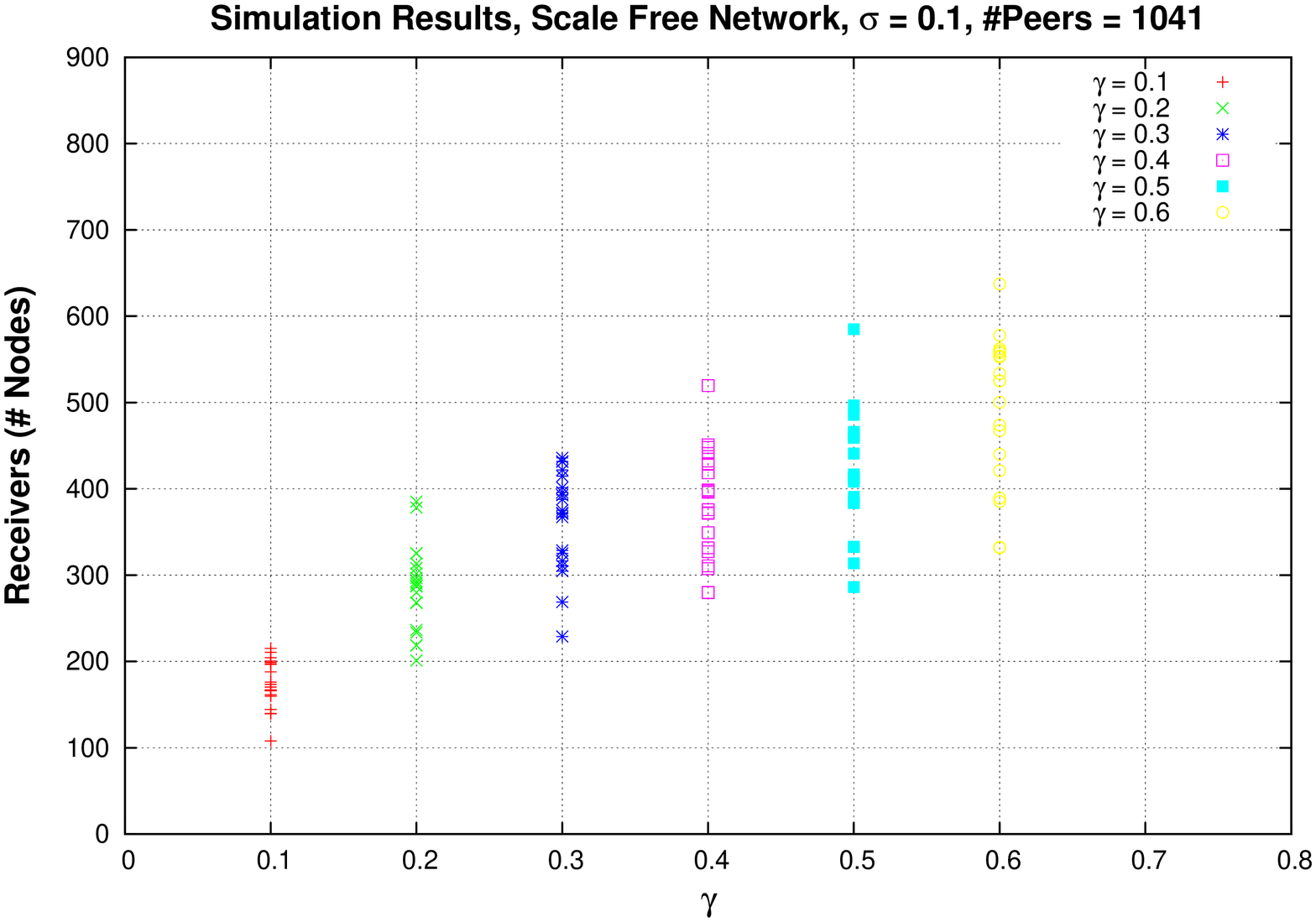}
   \caption{Model vs Simulation: scale free network, $a = 6, b = 1.4$, varying $\gamma$ above the phase transition. Number of receiving nodes obtained through simulation (the model returns an infinite sub-graph).}
   \label{fig:SF_confronto_s_beta1.4}
\end{figure}

\begin{figure}[t]
   \centering
   \includegraphics[angle=-90,width=\linewidth]{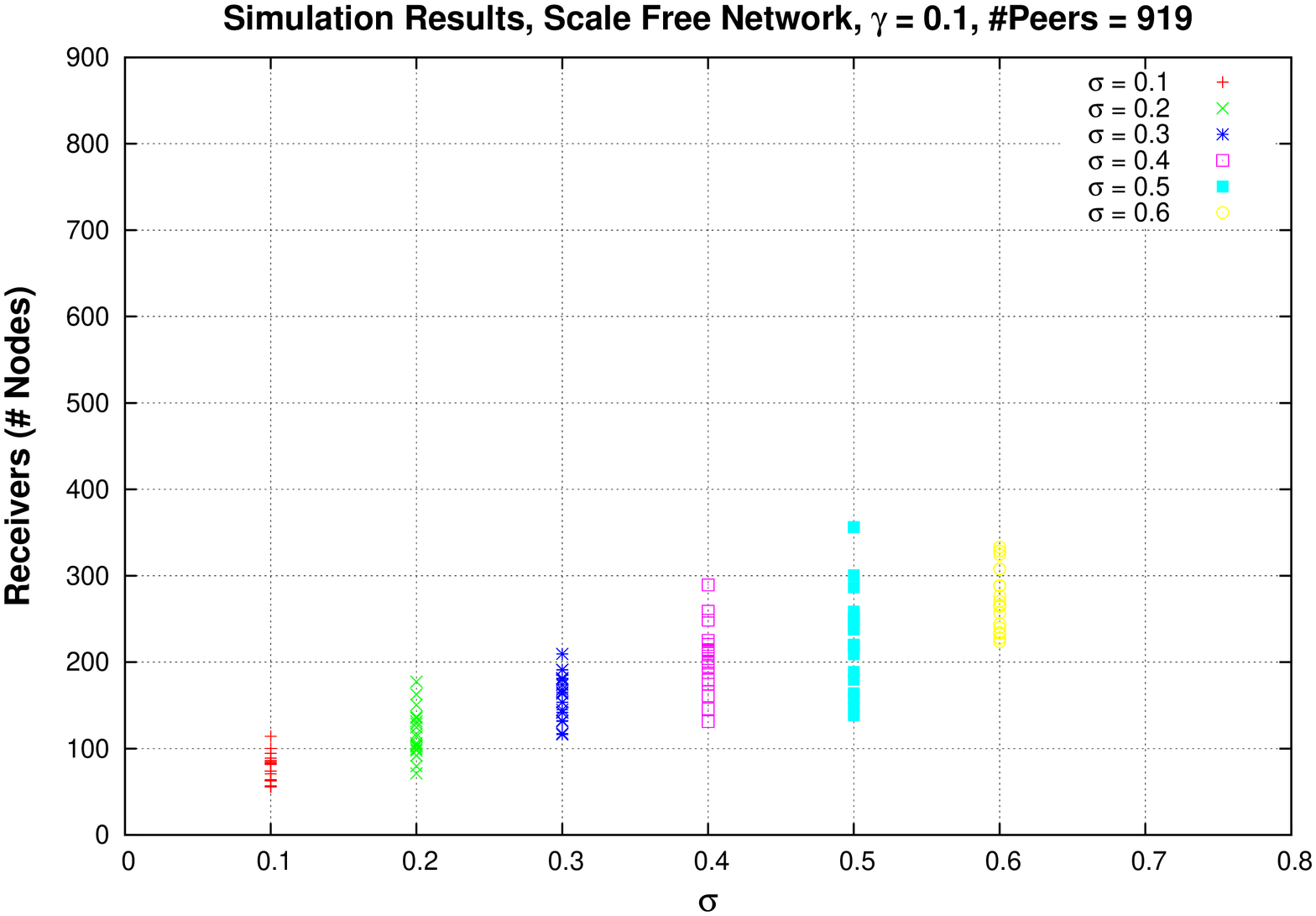}
   \caption{Model vs Simulation: scale free network, $a = 6, b = 1.5$, varying $\sigma$ above the phase transition. Number of receiving nodes obtained through simulation (the model returns an infinite sub-graph).}
   \label{fig:SF_confronto_g_beta1.5}
\end{figure}

\begin{figure}[t]
   \centering
   \includegraphics[angle=-90,width=\linewidth]{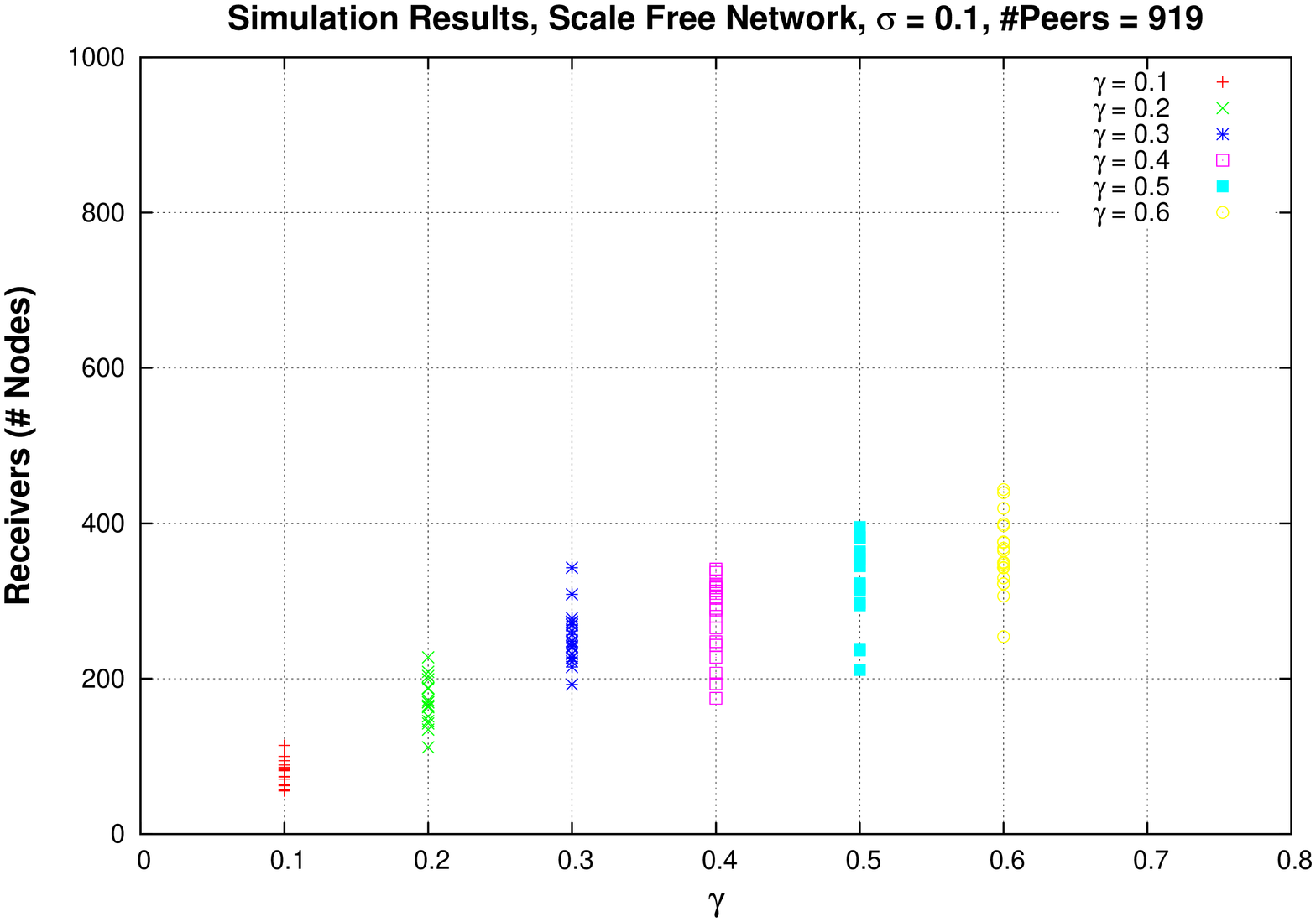}
   \caption{Model vs Simulation: scale free network, $a = 6, b = 1.5$, varying $\gamma$ above the phase transition. Number of receiving nodes obtained through simulation (the model returns an infinite sub-graph).}
   \label{fig:SF_confronto_s_beta1.5}
\end{figure}

To build scale-free networks, our simulator implements a construction method which has been proposed in \cite{Aiello00arandom}.
The interesting aspect of this algorithm is that it differs from other proposals, which build networks with a power law distribution by continuously adding novel nodes and edges, hence having networks that grow in time \cite{Barabasi2000}. Conversely, the method in \cite{Aiello00arandom} builds a network of fixed size, characterized by two parameters $a, b$. More specifically, 
the number of nodes $y$ which have a degree $x$ satisfies $\log{y} = a - b \log{x}$, i.e.~$y = \lfloor\frac{e^a}{x^b}\rfloor$. Thus, the total number of nodes of the generated network is
$$N = \sum_{x=1}^{\lfloor e^{\frac{a}{b}}\rfloor} \frac{e^a}{x^b},$$
being $\lfloor e^{\frac{a}{b}}\rfloor$ the maximum possible degree of the network, since it must be that $0 \leq \log{y} = a - b \log{x}$.
Once the number of nodes and their degrees have been determined, edges are randomly created among nodes until nodes reach their desired degrees.

Figure \ref{fig:fig_rete_Aiello} shows some examples of networks built with our simulator, implementing the construction method proposed in \cite{Aiello00arandom}. In particular, the chart reports, for three different settings of $a, b$, the number of nodes which have a given degree, in a log-log scale. It is possible to appreciate how such distributions are almost linear in a log-log scale, hence confirming they all follow some power law function.

As made above for random graphs, Figures \ref{fig:SF_confronto_g_above}, \ref{fig:SF_confronto_s_above} show results obtained in our simulations when we employ a scale-free network topology, with $\gamma=0.1$ (resp.~$\sigma=0.1$), while varying $\sigma$ (resp.~$\gamma$), above the phase transition. Again, based on the model an infinite number of receivers is reached (assuming a network of infinite size). From the simulations, a non-negligible portion of nodes is reached during the dissemination of events, that increases together with the $\gamma$ (resp.~$\sigma$) parameter. 
Indeed, it is interesting to observe that when $\gamma = 0.6$, $\sigma = 0.1$ almost all network peers receive the event during the dissemination, and thus, almost all subscribers receive the published events. In the scenarios reported in the pictures, in fact, we employed scale-free networks generated through the choice of $a=6$, $b=1$, resulting in networks composed of $2482$ nodes. In this case, simulation results provide average results above $2200$ nodes.
A similar behavior is obtained when $\sigma = 0.6$, $\gamma = 0.1$.
Again, this result is in accordance with the outcomes from the model, stating that an infinite number of nodes is reached with these settings.

Figures \ref{fig:SF_confronto_g_beta1.1}--\ref{fig:SF_confronto_s_beta1.5} show similar results for different networks settings. A significant portion of network nodes is reached, whose size increases together with the $\gamma, \sigma$ values. Again, all this confirms that the theoretical model is able to predict that a given event, published in the \ac{P2P}~publish-subscribe system, can percolate through the whole overlay.

\section{Conclusions}\label{sec:conc}

This paper analyzed the performance of an unstructured P2P overlay network that exploits a very simple dissemination strategy to build \ac{P2P} publish-subscribe systems. 
Results show that by tuning the gossip probability it is possible to spread contents through the overlay, without the need to resorting to sophisticated dissemination strategies built on top of costly structured distributed systems.
This is true when networks are large in size and the number of subscribers is not negligible.

In this work we focused on the network coverage. As concerns the communication overhead,
it is evident that the use of more costly solutions, such as centralized approaches or structured overlays, would provide better performances. In any case, the protocol limits the amount of messages sent in the network, since each node relays a given event only once. Hence, no duplicate transmissions occur on a link. Moreover, the low clustering guarantees that tree-like overlay are obtained, hence limiting the possibility that a peer receives multiple messages containing the same event. This is accomplished without the need (and the costs) of maintaining a structured overlay.

The mathematical framework proposed in the paper is quite general and can be exploited to model several types on unstructured overlays composing a \ac{P2P}~system. Focusing on the specific model for \ac{P2P}~publish-subscribe systems, there are several possible future works. For instance, the model could be extended to consider possible buffer overflows occurring when the event generation rate is higher than that which can be properly handled by peers in the overlay.

%
%

\end{document}